\documentclass[aps,prd,twocolumn,superscriptaddress]{revtex4}
\usepackage{epsfig,epsf}
\usepackage{amsmath}
\usepackage{amsthm}
\usepackage{amsfonts}
\usepackage{amssymb}
\usepackage{dsfont}
\usepackage{multirow}
\usepackage{appendix}
\usepackage{slashed}
\usepackage[active]{srcltx}
\usepackage{psfrag}

\begin{document}

\title{The Scalar Hexaquark $uuddss$: a Candidate to Dark Matter?}
\author{K.~Azizi}
\thanks{Corresponding author}
\affiliation{Department of Physics, University of Tehran, North Karegar Ave., Tehran
14395-547, Iran}
\affiliation{Department of Physics, Do\v{g}u\c{s} University, Acibadem-Kadik\"{o}y, 34722
Istanbul, Turkey}
\author{S.~S.~Agaev}
\affiliation{Institute for Physical Problems, Baku State University, Az--1148 Baku,
Azerbaijan}
\author{H.~Sundu}
\affiliation{Department of Physics, Kocaeli University, 41380 Izmit, Turkey}

\begin{abstract}
It is conventionally argued that Dark Matter (DM) has a non-baryonic nature,
but if we assume that DM was frozen out before primordial nucleosynthesis
and could not significantly impact primordial abundances this argument may
be evaded. Then a hypothetical $SU(3)$ flavor-singlet, highly symmetric,
deeply bound neutral scalar hexaquark $\mathrm{S}=uuddss$, which due to its
features has escaped from experimental detection so far, may be considered
as a candidate for a baryonic DM. In the present work we calculate the mass
and coupling constant of the scalar six-quark particle $\mathrm{S}$ by means
of the QCD sum rule method. Our predictions for its mass are $%
m_{S}=1180_{-26}^{+40}~\mathrm{MeV}$ ($m_s=95~\mathrm{MeV}$) and $%
\widetilde {m}_{S}=1239_{-28}^{+42}~\mathrm{MeV}$ ($m_s=128~\mathrm{MeV}$).
Although these values of mass would produce thermally the cosmological DM
abundance, existence of this state may contradict to stability of the oxygen
nuclei, which requires further thorough analysis.
\end{abstract}

\maketitle

\textit{\textbf{Introduction:}} From first days of the quark-parton model
and Quantum Chromodynamics (QCD), hadrons with unusual quantum numbers
and/or multiquark contents attracted interest of physicists. The
conventional hadrons have quark-antiquark or three-quark compositions. Their
masses and quantum parameters $J^{PC}$ are in accord with predictions of
this scheme and can be calculated using standard methods of particle
physics. Unusual or exotic hadrons is expected to be built of four and more
valence quarks or contain valence gluons. A main reason triggered intensive
investigations of four-quark states was a mass hierarchy inside the lowest
scalar multiplet, which found its explanation in the context of the
four-quark model suggested by R.~Jaffe \cite{Jaffe:1976ig}. Starting from
2003, i.e. from first observation of the exotic meson $X(3872)$ theoretical
and experimental investigations of tetra and pentaquarks became one of the
interesting and rapidly growing branches of high energy physics. Now,
valuable experimental information collected during past years, as well as
theoretical progress achieved to date, form two essential components of the
exotic hadrons physics \cite%
{Chen:2016qju,Chen:2016spr,Esposito:2016noz,Ali:2017jda,Olsen:2017bmm}.

Another interesting result about multiquark hadrons with far-reaching
consequences was obtained also by R.~Jaffe \cite{Jaffe:1976yi}. He
considered six-quark (dibaryon) states built of only light $u$, $d$, and $s$
quarks that belong to flavor group $SU_{f}(3)$. By combining the color and
spin of quarks and forming $SU_{cs}(6)$ "colorspin" group, Jaffe analyzed
its representations and found that dibaryons only from the singlet and octet
representations of $SU_{f}(3)$ may be light enough to be bound or resonant.
Among six-quark states from these two representations $SU_{f}(3)$ singlet
particles have zero spins. At the same time, all of singlet and octet
dibaryons are strange particles, therefore structures containing merely $u$
and $d$ quarks cannot be bound and stable. Using the MIT quark-bag model for
analysis, Jaffe predicted existence of a $\mathrm{H}$-dibaryon, i.e., of
flavor-singlet and\ neutral six-quark $uuddss$ bound state with
isospin-spin-parities $I(J^{P})=0(0^{+})$. This double-strange six-quark
structure with mass $2150\ \mathrm{MeV}$ lies $80~\mathrm{MeV}$ below the $%
2m_{\Lambda }=2230~\mathrm{MeV}\ $ threshold and is stable against strong
decays. It can decay through weak interactions, which means that mean
lifetime of $\mathrm{H}$-dibaryon, $\tau \approx 10^{-10}\mathrm{s}$, is
considerably longer than that of conventional mesons. The $\mathrm{H}$%
-dibaryon with the mass obeying this limit is loosely-bound state, a subject
to weak transformations.

The original work \cite{Jaffe:1976yi} was followed by numerous theoretical
investigations, in which various models and methods of particle physics were
used to calculate the $\mathrm{H}$-dibaryon's mass \cite%
{Liu:1982wg,Yost:1985mj,Oka:1983ku,Balachandran:1985fb,Oka:1986fr,Straub:1988mz,Koike:1989ak,Larin:1985yt,Kodama:1994np}%
. As usual, results of these studies are controversial: thus, calculations
in the framework of the corrected MIT bag model led to $m_{H}=2240~\mathrm{%
MeV}$ which is just above the $2m_{\Lambda }$ threshold \cite{Liu:1982wg},
whereas in a chiral model the authors \cite{Yost:1985mj} found $m_{H}=1130~%
\mathrm{MeV.}$ To analyze $\Lambda -\Lambda $ interaction and estimate $%
\Lambda \Lambda $ binding energy other quark models were invoked as well
\cite{Oka:1983ku,Oka:1986fr,Straub:1988mz,Koike:1989ak}. The $\mathrm{H}$%
-dibaryon's mass extracted from the QCD two-point sum rules is consistent
with the original result of Jaffe \cite{Larin:1985yt,Kodama:1994np}. In
fact, $m_{H}$ from Ref.\ \cite{Larin:1985yt} varies in limits $2.0-2.4~%
\mathrm{GeV}$ and within an accuracy of the sum rule method $\sim 20\%$
agrees with the result of the quark-bag model. Calculations in Ref.\ \cite%
{Kodama:1994np} also confirmed existence of a bound state lying $40~\mathrm{%
MeV}$ below the $2m_{\Lambda }$ threshold. The lattice simulations performed
in Ref.\ \cite{Iwasaki:1987db} led to conclusion that $m_{H}$ was below the $%
2m_{N}$ threshold $1880~\mathrm{MeV}$. In this paper the authors took into
account the stability conditions of the nucleus and extracted $m_{H}\approx
1850~\mathrm{MeV}$. The later lattice studies confirmed existence of a
bound-state $\mathrm{H}$-dibaryon, and predicted its binding energy $\approx
74.6~\mathrm{MeV}$ \cite{Beane:2012vq} and $(19\pm 10)~\mathrm{MeV}$ \cite%
{Francis:2018qch}, respectively. In the context of the holographic QCD $%
\mathrm{H}$-dibaryon was explored in Ref.\ \cite{Suganuma:2016lmp}, in which
its mass was estimated about $m_{H}=1.7~\mathrm{GeV}$.

The hexaquark $\mathrm{S}$ (except for original papers, hereafter we use a
hexaquark instead of a six-quark state, and denote it by $\mathrm{S}$) was
searched for by KTeV, Belle, and BaBar collaborations in exclusive $%
S\rightarrow \Lambda p\pi ^{-}$ and inclusive $\Upsilon (1S)$ and $\Upsilon
(2S)$ decays, in processes $\Upsilon (2S,\ 3S)\rightarrow S\overline{\Lambda
}\overline{\Lambda }$ \cite%
{AlaviHarati:1999ds,Carames:2013hla,Kim:2013vym,Echenard:2018rfe}. All these
experiments could not find evidence for the hexaquark $\mathrm{S}$ near the
threshold $2m_{\Lambda }$ and were able only to impose limits on its mass $%
m_{S}$ the latest being $m_{S}<2.05~\mathrm{GeV}$.

Recent activities around the $\mathrm{S}$ is inspired by renewed suggestions
to consider it as a possible candidate to dark matter \cite%
{Farrar:2003gh,Farrar:2003qy,Farrar:2017eqq,Farrar:2017ysn,Farrar:2018hac}.
In accordance with this scenario if $m_{S}\leq 2(m_{p}+m_{e})=1877.6~\mathrm{%
MeV}$ the hexaquark is absolutely stable, because all possible decay
channels of a free $\mathrm{S}$ is kinematically forbidden. For the $m_{S}$
obeying the inequality $m_{S}<m_{\Lambda }+m_{p}+m_{e}=2054.5~\mathrm{MeV}$
the hexaquark decays through a double-weak interaction, but even in this
case its lifetime could be comparable with the age of the Universe. The
lower bound of $m_{S}$ is determined by a stability of ordinary nuclei,
which are stable if $m_{S}>m_{p}+m_{n}+m_{e}-2E$, where $2E$ is a binding
energy of $p+n$. Then, for masses $1860<m_{S}<1880~\mathrm{MeV}$, which
assures a stability of the hexaquark and conventional nuclei, the $\mathrm{S}
$ can explain both the relic abundance of the DM in the Universe and
observed DM to ordinary matter ratio with less than $15\%$ uncertainty \cite%
{Farrar:2018hac}. But even $\mathrm{S}$ with the mass in the range $%
1.3\lesssim m_{S}\lesssim 2m_{p}$ and with radius $(1/6-1/4)r_{p}$, where $%
r_{p}\approx 0.86~\mathrm{fm}$ is the a proton radius, is consistent with
the stability of nuclei, with $\Lambda $ decays in double-strange
hypernuclei and experimental limits on existence of exotic isotopes of
helium and other nuclei \cite{Farrar:2003gh}. There are, however, objections
to this picture connected with a production process of hexaquarks in the
early-universe \cite{Kolb:2018bxv}, or with observed supernova explosion
\cite{McDermott:2018ofd}.

The hexaquark $\mathrm{S}$ as a candidate to DM was recently analyzed in
Ref.\ \cite{Gross:2018ivp} as well. In this work the mass of $\mathrm{S}$
was evaluated by modeling it as a bound state of scalar diquarks. Using the
effective Hamiltonian to describe dominant spin-spin interactions in
diquarks \cite{Esposito:2016noz}, the authors expressed $m_{S}$ in terms of
constituent diquark masses $m_{ij}$ and chomomagnetic couplings $k_{ij}$.
The masses of diquarks and chomomagnetic couplings may be extracted from
analysis of baryon spectroscopy. Alternatively, $k_{ij}$ can also be fixed
to reproduce masses of the light scalar mesons $f_{0}(500)$, $K^{\ast }(800)$
$f_{0}(980)$, and $a_{0}(980)$ interpreted as tetraquarks \cite%
{Maiani:2004uc}. It turns out that spin-spin couplings in tetraquarks are
about a factor of four larger compared to the spin-spin couplings in the
baryons. Because the hexaquark itself is an exotic six-quark meson for
calculation of $m_{S}$ it is reasonable to employ parameters estimated from
analysis of the light tetraquarks. Calculations carried out in Ref. \cite%
{Gross:2018ivp} predict $m_{S}\approx 1.2\ \mathrm{GeV}$ which reproduces
the cosmological DM abundance, but may contradict to stability of oxygen
nuclei.

\textit{\textbf{Calculations:}} In the present work we calculate the mass of
$\mathrm{S}$ by treating it as a bound state of three scalar diquarks. To
this end, we employe the QCD two-point sum rule approach, which is one of
the powerful nonperturbative methods to explore hadrons. As starting point,
the method uses the correlation function%
\begin{equation}
\Pi (p)=i\int d^{4}xe^{ipx}\langle 0|\mathcal{T}\{J(x)J^{\dagger
}(0)\}|0\rangle,  \label{eq:CorrF}
\end{equation}%
and extract from its analysis sum rules to compute spectroscopic parameters
of the hexaquark. The main ingredient of this analysis is the interpolating
current $J(x)$ which we choose it in the following form
\begin{eqnarray}
J(x) &=&\epsilon ^{abc}\left[ u^{T}(x)C\gamma _{5}d(x)\right] ^{a}\left[
u^{T}(x)C\gamma _{5}s(x)\right] ^{b}  \notag \\
&&\times \left[ d^{T}(x)C\gamma _{5}s(x)\right] ^{c},  \label{eq:Curr}
\end{eqnarray}%
where $[q^{T}C\gamma _{5}q^{\prime }]^{a}=\epsilon ^{amn}[q_{m}^{T}C\gamma
_{5}q_{n}^{\prime }]$ and $a,\ b,\ c,\ m$, $n$ are color indices with $C$
being the charge-conjugation operator.

As is seen, the hexaquark is composed of the scalar diquarks $[q^{T}C\gamma
_{5}q^{\prime }]^{a}$ in the color antitriplet and flavor antisymmetric
states. These diquarks are most attractive ones \cite{Jaffe:2004ph}, and
six-quark mesons composed of them should be lighter and more stable than
bound states of other two-quarks. Mathematical manipulations to derive sum
rules for the mass and coupling of the hexaquark are carried out in
accordance with standard prescriptions of the method. Thus, first we express
the correlation function $\Pi (p)$ in terms of the hexaquark's mass $m_{S}$
and coupling $f_{S}$, as well as its matrix element
\begin{equation}
\langle 0|J|S\rangle =m_{S}f_{S}.  \label{eq:ME}
\end{equation}%
Separating from each another the ground-state term and contributions due to
higher resonances and continuum states for $\Pi ^{\mathrm{Phys}}(p)$ we get
\begin{equation}
\Pi ^{\mathrm{Phys}}(p)=\frac{\langle 0|J|S(p)\rangle \langle
S(p)|J^{\dagger }|0\rangle }{m_{S}^{2}-p^{2}}+\cdots.  \label{eq:Phys1}
\end{equation}%
The expression of the matrix element (\ref{eq:ME}) allows us to rewrite $\Pi
^{\mathrm{Phys}}(p)$ in the form
\begin{equation}
\Pi ^{\mathrm{Phys}}(p)=\frac{m_{S}^{2}f_{S}^{2}}{m_{S}^{2}-p^{2}}+\cdots,
\label{eq:Phys2}
\end{equation}%
where dots denote contributions of higher resonances and continuum states.

To calculate the QCD or OPE side of the sum rules, we insert the current $%
J(x)$ to Eq.\ (\ref{eq:CorrF}), contract relevant quark fields and obtain $%
\Pi ^{\mathrm{OPE}}(p)$ in terms of the quark propagators:
\begin{eqnarray}
&&\Pi ^{\mathrm{OPE}}(p)=  \notag \\
&&i\delta ^{af}\delta ^{a^{\prime }f^{\prime }}\delta ^{bd}\delta
^{b^{\prime }d^{\prime }}\delta ^{ce}\delta ^{c^{\prime }e^{\prime }}\int
d^{4}xe^{ipx}\{\mathrm{Tr}[S_{d}^{ee^{\prime }}(x)  \notag \\
&&\times \gamma _{5}\widetilde{S}{}_{s}^{ff^{\prime }}(x)\gamma _{5}]\mathrm{%
Tr}[S_{u}^{aa^{\prime }}(x)\gamma _{5}\widetilde{S}{}_{d}^{bb^{\prime
}}(x)\gamma _{5}]  \notag \\
&&\times \mathrm{Tr}[S_{u}^{cc^{\prime }}(x)\gamma _{5}\widetilde{S}{}%
_{s}^{dd^{\prime }}(x)\gamma _{5}]\}+\mbox{511 similar terms},  \notag \\
&&  \label{eq:cont}
\end{eqnarray}%
where $\widetilde{S}(x)=CS^{T}(x)C$. To proceed, we employ the $x$-space
light-quark propagator
\begin{eqnarray}
&&S_{q}^{ab}(x)=i\frac{\slashed x}{2\pi ^{2}x^{4}}\delta _{ab}-\frac{m_{q}}{%
4\pi ^{2}x^{2}}\delta _{ab}-\frac{\langle \overline{q}q\rangle }{12}\left(
1-i\frac{m_{q}}{4}\slashed x\right) \delta _{ab}  \notag \\
&&-\frac{x^{2}}{192}\langle \overline{q}g_{s}\sigma Gq\rangle \left( 1-i%
\frac{m_{q}}{6}\slashed x\right) \delta _{ab}  \notag \\
&&-\frac{ig_{s}G_{ab}^{\mu \nu }}{32\pi ^{2}x^{2}}\left[ \slashed x\sigma
_{\mu \nu }+\sigma _{\mu \nu }\slashed x\right] -\frac{\slashed %
xx^{2}g_{s}^{2}}{7776}\langle \overline{q}q\rangle ^{2}\delta _{ab}  \notag
\\
&&-\frac{x^{4}\langle \overline{q}q\rangle \langle g_{s}^{2}G^{2}\rangle }{%
27648}\delta _{ab}+\frac{m_{q}g_{s}}{32\pi ^{2}}G_{ab}^{\mu \nu }\sigma
_{\mu \nu }\left[ \ln \left( \frac{-x^{2}\Lambda ^{2}}{4}\right) +2\gamma
_{E}\right]  \notag \\
&&+\cdots,  \label{eq:LQProp}
\end{eqnarray}%
where $q=u,~d$ or $s$, $\gamma _{E}\simeq 0.577$ is the Euler constant, and $%
\Lambda $ is a scale parameter. We also use the notations $G_{ab}^{\mu \nu
}\equiv G_{A}^{\mu \nu }t_{ab}^{A},\ A=1,2,\cdots, 8$, and $t^{A}=\lambda
^{A}/2$, with $\lambda ^{A}$ being the Gell-Mann matrices.

After inserting the light-quark propagators into Eq.\ (\ref{eq:cont}), we
get the correlation function $\Pi ^{\mathrm{OPE}}(p)$ in terms of QCD
degrees of freedom. The next step is to perform the resultant Fourier
integrals over four-$x$. Afterwards we equate the invariant amplitudes $\Pi
^{\mathrm{Phys}}(p^{2})$ and $\Pi ^{\mathrm{OPE}}(p^{2})$ to find the
desired sum rule in momentum space. We apply the Borel transformation to
both sides of the obtained sum rule to suppress contributions of the higher
resonances and continuum states, and using the quark-hadron duality
assumption, which is a quintessence of the sum rule method, perform the
continuum subtraction. An equality derived after these manipulations,
contains the mass and coupling constant of the particle $\mathrm{S}$. To
find the sum rules for $m_{S}$ and $f_{S}$ we need an extra expression which
can be obtained by acting $d/d\left( -1/M^{2}\right) $ to the first
equality. The sum rules for $m_{S}$ and $f_{S}$ derived by this way have
perturbative and nonperturbative components. The latter contains vacuum
condensates of various local quark, gluon, and mixed operators, which appear
after sandwiching relevant terms in $\Pi ^{\mathrm{OPE}}(p)$ between vacuum
states.

As the hexaquark is composed of six quarks, relevant computations are
lengthy and time consuming. In Appendix we explain some details of
calculations, and write down explicitly the Borel transformed and subtracted
invariant amplitude $\Pi ^{\mathrm{OPE}}(M^{2},s_{0})$ including
nonperturbative terms up to dimension ten. The full expression of $\Pi ^{%
\mathrm{OPE}}(M^{2},s_{0})$ contains terms up to dimension thirty, therefore
we refrain from providing them here. In numerical computations, we take into
account all these higher dimensional terms bearing in mind that they appear
due to the factorization hypothesis as product of basic condensates.

In analyses and computations, we utilize the quark, gluon and mixed
condensates
\begin{eqnarray}
&&\langle \bar{q}q\rangle =-(0.24\pm 0.01)^{3}~\mathrm{GeV}^{3},\ \langle
\bar{s}s\rangle =0.8\ \langle \bar{q}q\rangle ,  \notag \\
&&\langle \overline{q}g_{s}\sigma Gq\rangle =m_{0}^{2}\langle \overline{q}%
q\rangle ,\ \langle \overline{s}g_{s}\sigma Gs\rangle =m_{0}^{2}\langle \bar{%
s}s\rangle ,  \notag \\
&&m_{0}^{2}=(0.8\pm 0.1)~\mathrm{GeV}^{2}  \notag \\
&&\langle \frac{\alpha _{s}G^{2}}{\pi }\rangle =(0.012\pm 0.004)~\mathrm{GeV}%
^{4},  \label{eq:Parameters}
\end{eqnarray}%
which are determined at the scale $\mu =1~\mathrm{GeV}$. We work in the
approximation $m_{u}=m_{d}=0$, but keep a dependence on $m_{s}$. The
parameter $\Lambda $ is varied within the limits $(0.5,\ 1)~\mathrm{GeV}$. \

Since the $\mathrm{S}$ is tightly bound state with the radius much smaller
than usual hadrons, its parameters should be explored at relatively large
momentum scales $\mu $. Therefore, we calculate the mass and coupling of the
hexaquark at $\mu =2~\mathrm{GeV}$ which corresponds to the radius $0.1$ $%
\mathrm{fm}$. To reveal a sensitivity of $m_{S}$ and $f_{S}$ on the scale $%
\mu $, we evaluate the same parameters at $\mu =1~\mathrm{GeV}$ as well.

The mass of the strange quark $m_{s}=95_{-3}^{+9}~\mathrm{MeV}$ in the $%
\overline{MS}$ scheme and at the scale $\mu =2~\mathrm{GeV}$ can be found in
Ref.\ \cite{Tanabashi:2018oca}. We evolve the condensates (\ref%
{eq:Parameters}) to this scale and perform numerical computations. At the
scale $\mu =1~\mathrm{GeV}$ calculations are carried out by employing (\ref%
{eq:Parameters}) and the mass $m_{s}$ at this scale that differs from PDG
value by a factor $1.35$.

Another important problem is a proper choice for the the Borel $M^{2}$ and
continuum threshold $s_{0}$ parameters. First of them has been introduced
upon Borel transformation, the second one is necessary to separate the
ground-state and continuum contributions from each another, as we previously
have mentioned. These parameters are not arbitrary, but should meet the
well-known requirements. Thus, at maximum value of the Borel parameter the
pole contribution ($\mathrm{PC}$) should constitute a fixed part of the
correlation function, whereas at minimum of $M^{2}$ it must be a dominant
contribution. We define $\mathrm{PC}$ in the form
\begin{equation}
\mathrm{PC}=\frac{\Pi ^{\mathrm{OPE}}(M^{2},s_{0})}{\Pi ^{\mathrm{OPE}%
}(M^{2},\infty )},  \label{eq:PC}
\end{equation}%
and at $M_{\mathrm{max}}^{2}$ impose on it a restriction $\mathrm{PC}>0.2$,
which is usual for multiquark hadrons. The minimum of the Borel parameter $%
M^{2}$ is fixed from convergence of the sum rules, i.e. at $M_{\mathrm{min}%
}^{2}$ contribution of the last term (or a sum of last few terms) cannot
exceed, for example, $0.01$ part of the whole result. There is an another
restriction on the lower limit $M_{\mathrm{min}}^{2}$: at this $M^{2}$ the
perturbative contribution has to prevail over the nonperturbative one.

The sum rule predictions, in general, should not depend on the parameter $%
M^{2}$. But in real calculations $m_{S}$ and $f_{S}$ demonstrate
sensitiveness to the choice of $M^{2}$ and one should find a plateau where
this dependence is minimal. The continuum threshold parameter $s_{0}$
separates a ground-state contribution from the ones arising from higher
resonances and continuum states. Stated differently, $s_{0}$ should be below
the first excited state of the hexaquark $\mathrm{S}$. Parameters of
conventional hadrons' excited states are known either from experimental
measurements or from alternative theoretical studies. In the lack of similar
information for multiquark hadrons, one fixes $s_{0}$ to achieve a maximum
for $\mathrm{PC}$ ensuring, at the same time, fulfilments of other
constraints, and keeping under control a self-consistency of computations.
The self-consistent analysis implies that a gap between the mass $m_{S}$ and
the parameter $\sqrt{s_{0}}$ used for its extraction should be within
reasonable limits of a few hundred $~\mathrm{MeV}$.

Performed analysis allows us to determine the working regions
\begin{equation}
M^{2}\in \lbrack 1.3,\ 1.6]~\mathrm{GeV}^{2},\ s_{0}\in \lbrack 2.5,\ 2.9]~%
\mathrm{GeV}^{2},  \label{eq:Wind}
\end{equation}%
which obey all aforementioned restrictions.
\begin{figure}[h]
\includegraphics[width=8.8cm]{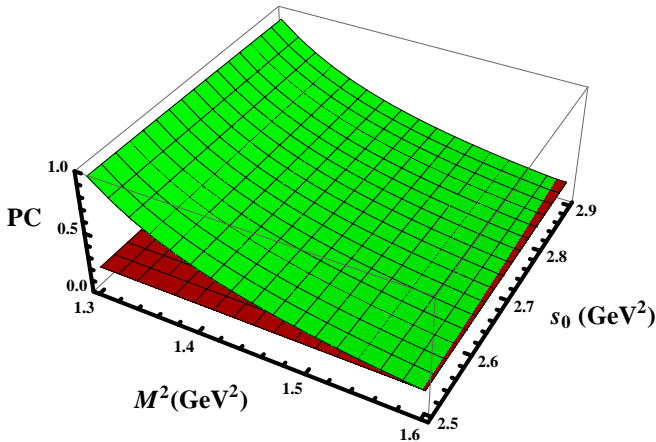}
\caption{Dependence of the pole contribution on $M^{2}$ and $s_{0}$. The
(red) surface $\mathrm{PC}=0.2$ is also shown.}
\label{fig:PC}
\end{figure}

\begin{figure}[h]
\includegraphics[width=8.8cm]{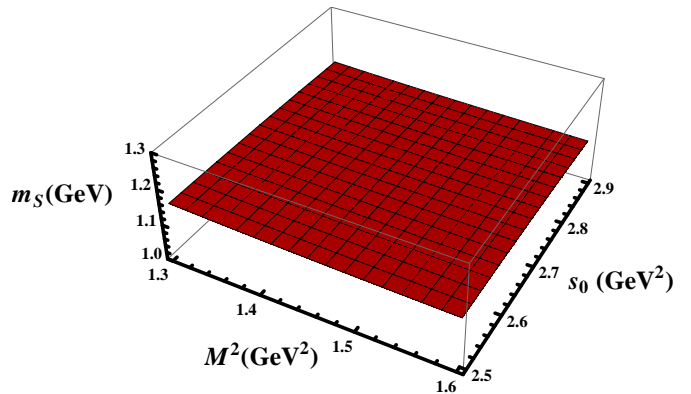}
\caption{ The mass of the hexaquark $\mathrm{S}$ as a function of the Borel
and continuum threshold parameters. The mass of the strange quark is $%
m_{s}=95~\mathrm{MeV}$.}
\label{fig:MassCoupl}
\end{figure}

In Fig.\ \ref{fig:PC} we plot the pole contribution, when $M^{2}$ and $s_{0}$
are varying within limits (\ref{eq:Wind}): at $M^{2}=1.3$ the pole
contribution is $0.9$, whereas at $M^{2}=1.6$ it becomes equal to $0.2$. The
predictions for the mass $m_{S}$ is pictured in Fig.\ \ref{fig:MassCoupl},
where a mild dependence on the parameters $M^{2}$ and $s_{0}$ is seen. The
results for the spectroscopic parameters of the hexaquark $\mathrm{S}$ read:
at the scale $\mu =2~\mathrm{GeV}$ (for $m_{s}=95~\mathrm{MeV}$)%
\begin{equation}
m_{S}=1180_{-26}^{+40}~\mathrm{MeV},\ f_{S}=8.56_{-0.26}^{+0.03}\times
10^{-6}~\mathrm{GeV}^{7},  \label{eq:CMass1}
\end{equation}%
and at the scale $\mu =1~\mathrm{GeV}$ (for $m_{s}=128~\mathrm{MeV}$)
\begin{equation}
\widetilde{m}_{S}=1239_{-28}^{+42}~\mathrm{MeV},\ \widetilde{f}%
_{S}=9.18_{-0.22}^{+0.03}\times 10^{-6}~\mathrm{GeV}^{7}.  \label{eq:CMass2}
\end{equation}%
Let us note that in computation of $\widetilde{m}_{S}$ and $\ \widetilde{f}%
_{S}$ the Borel parameter has been varied within the limits $M^{2}\in
\lbrack 1.34,\ 1.63]~\mathrm{GeV}^{2}$.

Theoretical uncertainties in the sum rule predictions (\ref{eq:CMass1}) and (%
\ref{eq:CMass2}) appear mainly due to working windows for the auxiliary
parameters $M^{2}$ and $s_{0}$, and the scale $\Lambda $. The ambiguities
connected with various vacuum condensates are numerically small. We do not
include into errors of the mass and coupling corrections generated by
different choices of the scale $\mu $, but keep ($m_{S},\ f_{S}$) and ($%
\widetilde{m}_{S},\ \widetilde{f}_{S}$) as two sets of parameters. Let us
note that variation of the mass $\Delta m_{S}(\mu )\approx 60~\mathrm{MeV}$
is comparable with other errors and does not exceed a few percent of $m_{S}$%
. It is not difficult to check also self-consistent character of obtained
results. Indeed, estimating $\sqrt{s_{0}}-m_{S}$ we get $[400,\ 525]~\mathrm{%
MeV}$, which can be accepted as a normal value for the mass difference
between the ground-state and first excited hexaquarks.

\textit{\textbf{Discussion and Conclusions:}} We have considered the spin-$0$%
, parity-even, highly symmetric $\mathrm{S}$-hexaquark of $uuddss$ with $Q=0$%
, $B=2$ and $S=-2$. Using the technique of QCD sum rule, we have found that
for the chosen interpolating current the intervals $M^{2}\in \lbrack 1.3,\
1.6]~\mathrm{GeV}^{2},\ s_{0}\in \lbrack 2.5,\ 2.9]~\mathrm{GeV}^{2}$ for
the auxiliary parameters fulfill the requirements of the method discussed
above. For these intervals, we could able to reach $[0.9-0.2]$ pole
contributions to the sum rules, and have extracted $m_{S}=1180_{-26}^{+40}~%
\mathrm{MeV}$ ($m_{s}=95~\mathrm{MeV}$) and $\widetilde{m}%
_{S}=1239_{-28}^{+42}~\mathrm{MeV}$ ($m_{s}=128~\mathrm{MeV}$) for the mass
of the $\mathrm{S}$-hexaquark. This range of the mass implies that the
hexaquark $S$ is an absolutely stable particle.

It is worth noting that in the context of the sum rule method the scalar
six-quark particle was investigated in Refs.\ \cite%
{Larin:1985yt,Kodama:1994np}. In Ref.\ \cite{Larin:1985yt}, the authors
explored different interpolating currents and carried out calculations by
taking into account nonperturbative terms up to $\langle \bar{q}q\rangle
^{4} $ and $m_{s}\langle \bar{q}q\rangle ^{5}$ orders. Using $m_{s}=0.2~%
\mathrm{GeV}$, and an approximation $\langle \bar{q}q\rangle =\langle \bar{s}%
s\rangle $, for the mass of the hexaquark they found $m_{S}=2.4~\mathrm{GeV}$%
. But, because of uncertainties of calculations, the authors could not
determine whether the mass of this particle lies above or below the $\Lambda
\Lambda $ threshold, i.e. whether it is stable or not. In Ref.\ \cite%
{Kodama:1994np} the two-point correlation function was found by employing
for the hexaquark a molecular type current. The authors took into account
only terms proportional to $\langle \bar{q}q\rangle ^{2}$ and $\langle \bar{q%
}q\rangle ^{4}$, and neglected the gluon and mixed condensates. Prediction
in Ref.\ \cite{Kodama:1994np} was made using the strange quark mass $%
m_{s}=150~\mathrm{MeV}$; it was found that the mass of the hexaquark is $%
m_{S}\approx 2.19~\mathrm{GeV}$ which corresponds to a bound state $40~%
\mathrm{MeV}$ below the $\Lambda \Lambda $ threshold.

The accuracy of calculations performed in the present work considerably
exceeds the accuracy of previous investigations. In this paper we have
included into analysis not only the quark, but also the gluon and mixed
condensates. In computations we took into account nonperturbative terms up
to dimension thirty, and for the mass of the strange quark used its
contemporary value $m_{s}=95_{-3}^{+9}~\mathrm{MeV}$. Therefore, our result
for the mass of $\mathrm{S}$-hexaquark differs from estimations of Refs.\
\cite{Larin:1985yt,Kodama:1994np}, but is in accord with output of the
chiral model \cite{Yost:1985mj}. Our prediction for $m_{S}$ (and $\widetilde{%
m}_{S})$ almost coincides with one made recently in Ref.\ \cite%
{Gross:2018ivp}, in which it was obtained by modeling the $\mathrm{S}$%
-hexaquark as a bound state of scalar diquarks and using recent progress in
theoretical and experimental physics of multiquark hadrons.

There are a lot of constraints on the mass and radius of the hexaquark as a
candidate to the DM. They stem from analyses of the different production
modes of $\mathrm{S}$-hexaquark in high energy hadron and $e^{-}e^{+}$
collisions, from its interactions with ordinary baryons, and from analysis
of cosmological processes \cite{Farrar:2017eqq,Farrar:2017ysn}. As we have
noted above, accelerator experiments could not observe the hexaquark $%
\mathrm{S}$ near the threshold $2m_{\Lambda }$ and only imposed limits on
its mass $m_{S}<2.05~\mathrm{GeV}$. There are some reasons that make
problematic detection of $\mathrm{S}$-hexaquark at colliders. In fact, the
hexaquark is neutral and compact particle and its mass is close to the mass
of neutron, which makes difficult to separate relevant signals from ones
generated by neutron. Additionally, as a flavor-singlet particle, the $%
\mathrm{S}$-hexaquark presumably does not couple to a hadronic content of
photon, pion and other flavor-nonsinglet mesons, or such interactions are
very weak. It is possible that due to these features it escapes detection in
hadron collisions. Strategies for discovering a stable hexaquark in various
hadronic processes and relevant problems were discussed in Ref.\ \cite%
{Farrar:2017eqq}. In accordance with estimates of this work, within existing
experimental datasets should be a few hundred events with $\mathrm{S}$ or
its antiparticle $\overline{\mathrm{S}}$. For creation of the $\mathrm{S}$%
-hexaquark $e^{-}e^{+}$ collisions may be more promising than hadronic
processes, because flavor-singlet multi-gluon states copiously produced in
these collisions eventually may transform to $\mathrm{S}$ particles \cite%
{Farrar:2017ysn}. In any case, the constraint $m_{S}<2.05~\mathrm{GeV}$
extracted from collider processes does not contradict to existence of a
stable particle with the mass $m_{S}\approx 1.2~\mathrm{GeV}$.

As a particle carrying the baryon number $B=2$, the $\mathrm{S}$-hexaquark
interacts with other baryons and nuclei, and these processes might be
utilized to detect it. Parameters of such interactions have to be calculated
using perturbative or nonperturbative methods of QCD, and confronted with
available experimental data. There are some active experiments designed for
direct detection of DM scattering on target materials of detectors \cite%
{Aprile:2019xxb,Aprile:2019jmx,Agnese:2018gze}. Recently the XENON and
SuperCDMS collaborations reported about constraints on light DM-nucleon
scattering cross-section $\sigma _{DMN}$ extracted from their studies \cite%
{Aprile:2019jmx,Agnese:2018gze}. In accordance with Ref.\ \cite%
{Aprile:2019jmx}, for DM particles with the mass $\sim 1\ \mathrm{GeV}$ and
a spin-independent $DM-N$ interaction the cross-section $\sigma _{DMN}$ is $%
\sigma _{DMN}\approx 10^{-38}~\mathrm{cm}^{2}$; and $\sigma _{DMp}\approx
10^{-32}~\mathrm{cm}^{2}$ and  $\sigma _{DMn}\approx 10^{-33}~\mathrm{cm}^{2}$
if this interaction is spin dependent (see, Fig.\ 5 in Ref.\ \cite%
{Aprile:2019jmx} ). The information provided by the SuperCDMS (see, Fig.\ 13
of \cite{Agnese:2018gze}) allows us to estimate $\sigma _{DMN}\sim 10^{-36}~%
\mathrm{cm}^{2}$. The limit on DM-proton cross-section $\sigma _{DMp}\leq
0.6\times 10^{-30}~\mathrm{cm}^{2}$ was extracted also in Ref.\ \cite%
{Mahdawi:2017cxz}, in which the authors relied on XDC data \cite%
{McCammon:2002gb}. Results of other experiments devoted to DM-nucleon
scattering can be found in Refs.\ \cite%
{Aprile:2019jmx,Agnese:2018gze,Mahdawi:2017cxz}.

The theoretical investigation of the hexaquark-nucleon scattering
cross-section $\sigma _{NS}$ should be performed in the context of QCD,
i.e., in a framework of the Standard Model (SM). In this aspect, this task
differs considerably from the  situation in theories, where DM particle and a
mediator of DM-nucleon interaction are introduced at expense of various
extensions of SM \cite{Bertone:2004pz}. Because the hexaquark is the
particle composed of the conventional quarks, a mediator of $\mathrm{S}-N$
interaction also should belong to ordinary hadron spectroscopy. It should be
neutral flavor-singlet scalar particle with a mass $<1~\mathrm{GeV}$: There
are few candidates to play a role of such mediator. The scalar mesons $%
\sigma $ [in new classification $f_{0}(500)$ ], $f_{0}(980)$, and a scalar
glueball $G$ may couple to both $\mathrm{S}$ and baryons, and carry this
interaction. The $\sigma $ and $f_{0}(980)$ are singlets from the lowest
nonet of scalar mesons. Difficulties in interpretation of these mesons as $q%
\overline{q}$ states inspired suggestion about their diquark-antidiquark
nature \cite{Jaffe:1976ig}. Within this paradigm problems with low masses,
and a mass hierarchy inside the light nonet seem found their solutions. The
current status of relevant theoretical studies can be found in Refs.\ \cite%
{Kim:2017yvd,Agaev:2017cfz,Agaev:2018sco}. The scalar glueball $G$, in
accordance with various estimations, has the mass $\gtrsim 1~\mathrm{GeV}$,
but due to mixing with quark component would appear as a part of $\sigma $
and $f_{0}(980)$ mesons \cite{Ochs:2013gi}.

Here, for the sake of concreteness, we analyze only $\sigma $-exchange
processes. Then, to evaluate the cross-section $\sigma _{NS}$ one has to
compute strong couplings corresponding to vertices $NN\sigma $ and $\mathrm{%
SS}\sigma $: This is necessary to evaluate $\sigma _{NS}$ using one of QCD
approaches. The $NN\sigma $ coupling, actually, is known, and was calculated
in the context of the sum rule method in Refs.\ \cite%
{Kisslinger:1999jk,Erkol:2005jz}. In these articles the meson $\sigma $ was
treated either as mixed $q\overline{q}+G$ or pure $q\overline{q}$ states.
The coupling $NN\sigma $, where $\sigma $ is the diquark-antidiquark exotic
meson or admixture of four-quark and glueball components, as well as $%
\mathrm{SS}\sigma $ were not calculated using available methods of QCD. It
is clear, that the hexaquark $\mathrm{S}$ is self-interacting particle,
which takes place through the same $\sigma $-exchange mechanism. In other
words, the $\mathrm{S}$-hexaquark as the DM particle belongs to class of
self-interacting DM models. The cross-section $\sigma _{SS}$ for elastic $%
\mathrm{S-S}$ interaction can be computed using the strong coupling of the
vertex $\mathrm{SS}\sigma $. It is worth noting that $\sigma _{NS}$ was
evaluated in Ref.\ \cite{Farrar:2003gh} by modeling $\mathrm{S}-N$
interaction via one-boson exchange Yukawa potential. The result obtained
there put on $\sigma _{NS}$ the limit $\sigma _{NS}\lesssim 10^{-3}~\mathrm{%
mb}$ or $\lesssim 10^{-30}~\mathrm{cm}^{2}$. At high velocities of the
interacting particles $v\approx c$ , naive estimates for $\sigma _{NS}$ and $%
\sigma _{SS}$ led to constraints $\sigma _{NS}\geq 0.25\sigma _{NN}\approx 5~%
\mathrm{mb}$ and $\sigma _{SS}\geq 0.25\sigma _{NS}\approx 1.25~\mathrm{mb}$
\cite{Farrar:2017ysn}, respectively. The restrictions on $\sigma _{DMp}$ for
spin-dependent interactions $10^{-27}~\mathrm{cm}^{2}<\sigma _{DMp}<10^{-24}~%
\mathrm{cm}^{2}$ were obtained in Ref.\ \cite{Hooper:2018bfw}. A strong
theoretical analysis, in accordance with the scheme outlined above, is
required to make a reliable prediction for $\sigma _{NS}$ compatible or not
with existing experimental limits.

Besides direct detection experiments, there are outer space cosmic ray (CR)
experiments, results of which can be utilized to explore the Dark Matter
through its decays to SM particles. Such investigations may be helpful also
for studying matter-antimatter asymmetry in the Universe. The finished
satellite-based PAMELA and ongoing AMS-02 experiments provided information
useful for these purposes. The PAMELA was constructed to detect galactic
CRs, and mainly their positron and antiproton components \cite%
{Adriani:2017bfx}. Measurements revealed a rise in the positron-electron
ratio $e^{+}/(e^{-}+e^{+})$ at energies above $10~\mathrm{GeV}$. The
abundance of positrons in CRs was confirmed by AMS-02 up to energies $350~%
\mathrm{GeV}$ \cite{Aguilar:2013qda} and later till $500~\mathrm{GeV}$. The
similar enhancement was observed in antiproton to proton ratio $\overline{p}%
/p$, as well \cite{Adriani:2010rc,Aguilar:2016kjl}. A standard scenario
implies that antiparticles are produced due to inelastic interactions of CR
nuclei with particles of the interstellar gas. But rates of such processes
are small, therefore deviation of these ratios from expected small values
may be interpreted in favor of DM decaying to SM particles. Collected data
on antiparticle anomaly in the Universe provide valuable information to
verify different DM models, and impose constraints on DM-SM interactions
\cite{Bertone:2004pz}. Alternatively, the observed abundance of high-energy
positrons in CRs may be connected with accelerating effects of nearby
sources, such as a pulsar, supernova remnants \cite%
{Aharonian:1995zz,Blasi:2009hv,Ahlers:2009ae}. This mechanism implies some
anisotropy in detected antiparticle fluxes, whereas data are consistent with
their isotropic distributions: this is significant obstacle in attempts to
explain antiparticle excess in CRs using local sources.

The hexaquark $\mathrm{S}$ produces leptons in the $\mathrm{S}\overline{%
\mathrm{S}}$ annihilation and inelastic interactions with particles of the
interstellar gas. Productions of various groups of $\sigma $, $f_{0}(980)$, $%
\pi $, and $K$ mesons would be main channels at low energy annihilations. At
high energies these light mesons should be accompanied by numerous baryons.
Dominant decay modes of the mesons $\sigma $, $f_{0}(980)$, $\pi $, and $K$,
and relevant branching ratios are well known \cite{Tanabashi:2018oca}, which
can be employed to estimate production rates of electrons and positrons to
account for discussed effects.

Important restrictions on the mass of the hexaquark arise from observed
cosmological abundance of the DM, and DM to ordinary matter ratio in the
Universe. The concept of DM is necessary (excluding models with modified
theory of gravity) to account for observed astrophysical effects, such as
rotation curves of galaxies, gravitational lensing, and other phenomena.
Direct evidence for the existence of DM came from analysis of bullet
clusters \cite{Clowe:2003tk}. All these phenomena can be explained within
the concept of DM provided DM particles are stable. The hexaquark with mass $%
m_{S}\approx 1.2~\mathrm{GeV}$ reproduces observed DM abundance and
DM/matter ratio, while a larger $m_{S}$ gives a smaller relic abundance \cite%
{Gross:2018ivp}. The $\mathrm{S}$-hexaquark is the stable particle, and its
self-interaction and elastic scattering on ordinary matter does not reduce
the total mass of DM in some location, which is crucial to describe
aforementioned phenomena. Of course, interaction of $\mathrm{S}$ with baryon
and photon fluids may alter matter power spectrum and the cosmic microwave
background, but they are not strong enough to generate significant effects
\cite{Gross:2018ivp}. Contrary, the $\mathrm{S}$-baryon (i.e., DM-galactic
gas) interactions may produce a DM disk embedded within the spherical
galactic halo, lead to co-rotation of DM with the gas and forming DM density
structure similar to that of the gas \cite{Farrar:2017ysn}. Parameters of
the DM disk in a galaxy, it thickness, for example, depend on the mass of DM
and the cross-section $\sigma _{DMN}$ \cite{Wadekar:2019xnf}.

Another constraint on $m_{S}$ is connected with stability of conventional
nuclei. The reason is that very small mass of the hexaquark may contradict
to stability of existing nuclei, because for small $m_{S}$ nucleons inside
nuclei would bind to the $\mathrm{S}$ state faster than what is allowed by
observed stability of these nuclei. This process runs through double-weak
production of the off-shell $\Lambda ^{\ast }$ baryons by a pair of nucleons
$pn$, $pp$ or $nn$. Because our estimate for the mass of the hexaquark is $%
m_{S}\approx 1.2\ \mathrm{GeV}$, the main sources of the virtual $\Lambda
^{\ast }$ baryons are the weak decays $p\rightarrow \Lambda ^{\ast }\pi ^{+}$%
, $n\rightarrow \Lambda ^{\ast }\pi ^{0}$, and an internal conversion $%
(udd)\rightarrow (uds)$. Generated by this way virtual $\Lambda ^{\ast }$s
afterwards through the strong-interaction process $\Lambda ^{\ast }\Lambda
^{\ast }\rightarrow S$ form the hexaquark $\mathrm{S}$.

The matrix element of the reaction $NN\rightarrow SX$ can be written as a
product of the amplitude for the nucleons' double-weak transitions to a pair
of virtual $\Lambda ^{\ast }$s, and matrix element for creation of the $%
\mathrm{S}$ from the $\Lambda ^{\ast }$s \cite{Farrar:2003qy}
\begin{eqnarray}
\mathcal{M}(NN &\rightarrow &SX)\approx \mathcal{M}(NN\rightarrow \Lambda
^{\ast }\Lambda ^{\ast }X)\mathcal{M}(\Lambda ^{\ast }\Lambda ^{\ast
}\rightarrow S).  \notag \\
&&  \label{eq:Ampl}
\end{eqnarray}%
Then lifetime of the nucleus $\mathcal{N}$ decaying to $\mathcal{N}^{\prime
} $ and the hexaquark is%
\begin{equation}
\tau (\mathcal{N}\rightarrow \mathcal{N}^{\prime }SX)\simeq \frac{3\mathrm{yr%
}}{|\mathcal{M}(\Lambda ^{\ast }\Lambda ^{\ast }\rightarrow S)|^{2}},
\label{eq:lifetime1}
\end{equation}%
which should be confronted with the Super-Kamiokande (SK) limit for the
oxygen nuclei
\begin{equation}
\tau ({}^{16}O_{8}\rightarrow \mathcal{N}^{\prime }SX)\gtrsim 10^{26}\mathrm{%
yr}.  \label{eq:lifetime2}
\end{equation}%
Equation (\ref{eq:lifetime1}) is rather rough estimate for $\tau $, which is
seen from treatment, for instance, of the matrix element $|\mathcal{M}%
(\Lambda \rightarrow N)|^{2}$ used to derive it \cite{Farrar:2003qy}). This
matrix element was calculated there in the harmonic oscillator model, and
was also inferred from phenomenological analysis: obtained results differ
from each other by approximately $10$ times. The prediction (\ref%
{eq:lifetime1}) was made by employing an average value of $|\mathcal{M}%
(\Lambda \rightarrow N)|^{2}$.

The situation with $|\mathcal{M}(\Lambda ^{\ast }\Lambda ^{\ast }\rightarrow
S)|^{2}$ is even worst than in the previous case. The matrix element $|%
\mathcal{M}(\Lambda ^{\ast }\Lambda ^{\ast }\rightarrow S)|^{2}$ describes
the strong process $\Lambda ^{\ast }\Lambda ^{\ast }\rightarrow S$ and plays
a crucial role in theoretical estimations of the $^{16}O_{8}$ lifetime. In
Ref.\ \cite{Farrar:2003qy} the $\mathcal{M}(\Lambda ^{\ast }\Lambda ^{\ast
}\rightarrow S)$ was calculated as overlap integral of the final hexaquark
and initial $\Lambda ^{\ast }\Lambda ^{\ast }$ baryons's wave functions, the
latter being factored into wave functions of the two $\Lambda ^{\ast }$
baryons, and a wave function of two nucleons inside nucleus. The $\Lambda
^{\ast }$ baryon and hexaquark $\mathrm{S}$ wave functions were written down
using the Isgur-Karl (IK) nonrelativistic harmonic oscillator quark model
\cite{Isgur:1978wd} and its generalization to six-quark system \cite%
{Farrar:2003qy}. These functions depend on parameters $\alpha _{B(S)}=1/%
\sqrt{\langle r_{B(S)}^{2}\rangle }$, where $\langle r_{B}^{2}\rangle $ and $%
\langle r_{S}^{2}\rangle $ are mean charge radii of the $\Lambda ^{\ast }$
baryon and hexaquark $\mathrm{S}$, respectively. The features of the two
nucleons inside nucleus were modeled in Ref.\ \cite{Farrar:2003qy} using
Brueckner-Bethe-Goldstone (BBG) wave function. A wide class of the
two-nucleon wave functions including the BBG, the Miller-Spencer wave
functions, and ones extracted from results of Ref.\ \cite{Lonardoni:2017egu}%
, was employed in Ref.\ \cite{Gross:2018ivp} to study the stability of the
oxygen nucleus.

The IK wave functions used to model the $\Lambda ^{\ast }$ baryons and the
hexaquark $\mathrm{S}$ suffer from serious drawbacks. Thus, the parameter $%
\alpha _{B}=0.406~\mathrm{fm}^{-1}$ necessary to describe the mass splitting
of lowest lying $\frac{1}{2}^{+}$ and $\frac{3}{2}^{+}$ baryons corresponds
to radius of the proton $0.49~\mathrm{fm}$, whereas to reproduce the
experimental value $0.86~\mathrm{fm}$ one needs $\alpha _{B}=0.221~\mathrm{fm%
}^{-1}$. In other words, the IK functions could not explain simultaneously
the mass splitting of positive parity baryons and the proton radius. Usage
of such nonrelativistic wave function to describe the $\Lambda ^{\ast }$
baryon, moreover an attempt to generalize and apply it to the relativistic
multiquark system like the hexaquark is, at least, questionable. Existing
problems of the IK model were emphasized already in the original paper \cite%
{Farrar:2003qy} and repeated in Ref.\ \cite{Gross:2018ivp}, nevertheless in
both of them these wave functions were employed to carry out numerical
computations.

Calculations demonstrate that the matrix element $|\mathcal{M}(\Lambda
^{\ast }\Lambda ^{\ast }\rightarrow S)|^{2}$ is highly sensitive to the
choice of the parameters $\alpha _{B}$ and $\alpha _{S}$, and is suppressed
if the hexaquark $\mathrm{S}$ has a small radius $r_{S}/r_{N}\ll 1$, where $%
r_{N}$ is the nucleon radius. The radius of the hexaquark was estimated in
Ref.\ \cite{Farrar:2018hac} as $0.1\leq r_{S}\leq 0.3\ \mathrm{fm}$ which
implies $r_{S}/r_{N}=0.11-0.34$. Then, for example, for $r_{S}\approx 0.13~%
\mathrm{fm}$ and, as a result, for the ratio $r_{N}/r_{S}\approx 6.6$ the
BBG wave function with the core radius $c\simeq 0.4~\mathrm{fm}$, and $%
\alpha _{B}=0.406~\mathrm{fm}^{-1}$ satisfies the constraint $|\mathcal{M}%
(\Lambda ^{\ast }\Lambda ^{\ast }\rightarrow S)|^{2}\lesssim 10^{-25}$
necessary to evade the SK bound (see, Fig.\ 1 in Ref. \cite{Farrar:2003qy}).
At the same parameters of the hexaquark, but for $\alpha _{B}=0.221~\mathrm{%
fm}^{-1}$ the core radius in the BBG model should be larger to achieve
required suppression of the matrix element $|\mathcal{M}(\Lambda ^{\ast
}\Lambda ^{\ast }\rightarrow S)|^{2}$.

To update predictions for $|\mathcal{M}(\Lambda ^{\ast }\Lambda ^{\ast
}\rightarrow S)|^{2}$, the authors in Ref.\ \cite{Gross:2018ivp} used the
new two-nucleon wave functions. The latter were extracted by utilizing an
information on the two-nucleon point density inside nuclei $\rho _{NN}(r)$
\cite{Lonardoni:2017egu}. The original calculations of $\rho _{NN}(r)$ were
performed by employing the nonrelativistic Hamiltonian, where the
phenomenological $NN$ potential includes electromagnetic and
one-pion-exchange terms, and also contains phenomenological contributions to
reproduce nucleon-nucleon elastic scattering. The new wave functions, of
course, present a more detailed picture of a nucleus, but are
nonrelativistic quantities and do not take into account dynamical effects of
quark-gluon interactions in nuclei. The updated results for $|\mathcal{M}%
(\Lambda ^{\ast }\Lambda ^{\ast }\rightarrow S)|^{2}$ are presented in Fig.\
5 of Ref.\ \cite{Gross:2018ivp} as a function of the ratio $r_{S}/r_{N}$.
Unfortunately, the authors did not show the region $r_{S}/r_{N}=0.11-0.2$,
in which the restriction on the matrix element $|\mathcal{M}(\Lambda ^{\ast
}\Lambda ^{\ast }\rightarrow S)|^{2}\lesssim 10^{-25}$ may be satisfied.

As it has been just emphasized above, $|\mathcal{M}(\Lambda ^{\ast }\Lambda
^{\ast }\rightarrow S)|^{2}$ critically depends on a behavior of the
relevant wave functions at small inter-nucleon distances $r\lesssim 1~%
\mathrm{fm}$. At these distances nucleons are not the suitable degrees of
freedom, and quark-gluon content of the nucleons becomes essential to
describe correctly processes inside nuclei. Neither the Isgur-Karl type wave
functions of the $\Lambda ^{\ast }$ baryon and $\mathrm{S}$-hexaquark, nor
the two-nucleon wave functions discussed till now contain detailed
information on relativistic and nonperturbative interactions of quarks and
gluons at distances $r\lesssim 1~\mathrm{fm}$ and high densities. Therefore,
the predictions for $|\mathcal{M}(\Lambda ^{\ast }\Lambda ^{\ast
}\rightarrow S)|^{2}$ made in Refs.\ \cite{Farrar:2003qy,Gross:2018ivp}
cannot be considered as credible ones and used to confirm or exclude
existence of the hexaquark $\mathrm{S}$. Only after thorough exploration of
aforementioned problems, we can get serious estimate for $|\mathcal{M}%
(\Lambda ^{\ast }\Lambda ^{\ast }\rightarrow S)|^{2}$ and see whether the
requirement $r_{S}<<r_{N}$--necessary for stability of $^{16}O_{8}$ in the
present picture--survives or not. Then we will be able to answer the
question moved to the title of the present article as well.

\appendix*

\section{Details of calculations and the correlation function $\Pi ^{\mathrm{%
OPE}}(M^{2},s_{0})$}

\renewcommand{\theequation}{\Alph{section}.\arabic{equation}} \label{sec:App}

In this appendix we present some details of calculations, which are
necessary to derive the sum rules for the mass and coupling of the
hexaquark. It is evident that a main problem is calculation of the QCD side
of the sum rules $\Pi ^{\mathrm{OPE}}(p)$. Using explicit expressions for
the light quarks propagators and inserting relevant ones into Eq.\ (\ref%
{eq:cont}), we get the Fourier integrals of following types:
\begin{eqnarray}
T[l,m] &=&\int d^{4}xe^{ipx}\frac{[1,x_{\mu },x_{\mu }x_{\nu
},...](x^{2})^{l}\left[ \ln \left( \frac{-x^{2}\Lambda ^{2}}{4}\right) %
\right] ^{m}}{(x^{2})^{n}}.  \notag \\
&&
\end{eqnarray}%
For simplicity, let us consider the case $l=0$ and $m=0$. After a Wick
rotation in the Euclidean space for $T[0,0]\equiv T$, one finds
\begin{equation}
T=-i(-1)^{n}\int d^{4}x_{E}\frac{e^{-ip_{E}x_{E}}}{(x_{E}^{2})^{n}}.
\end{equation}%
By applying the Schwinger parametrization
\begin{equation}
\frac{1}{A^{n}}=\frac{1}{\Gamma (n)}\int_{0}^{\infty
}dt~t^{n-1}e^{-tA},~~~~~A>0,
\end{equation}%
it is not difficult to recast $T$ into the form
\begin{equation}
T=-i(-1)^{n}\frac{1}{\Gamma (n)}\int_{0}^{\infty }dt~t^{n-1}\int
d^{4}x_{E}~e^{-ip_{E}x_{E}}e^{-tx_{E}^{2}}.  \notag \\
\end{equation}%
By performing the resultant Gaussian integral over four-$x$, we obtain
\begin{equation}
T=-i(-1)^{n}\frac{\pi ^{2}}{\Gamma (n)}\int_{0}^{\infty
}dt~t^{n-3}e^{-p_{E}^{2}/4t}.
\end{equation}%
The next step is to apply the Borel transformation with respect to $%
p_{E}^{2} $ to suppress contributions of the higher resonances and continuum
states. To this end, we utilize the formula
\begin{equation}
\mathcal{B}_{M}e^{-p_{E}^{2}/4t}=\delta \left( \frac{1}{M^{2}}-\frac{1}{4t}%
\right) ,
\end{equation}%
which leads to
\begin{equation}
\mathcal{B}_{M}T=-i(-1)^{n}\frac{\pi ^{2}}{\Gamma (n)}\int_{0}^{\infty
}dt~t^{n-3}\delta \left( \frac{1}{M^{2}}-\frac{1}{4t}\right) .
\end{equation}%
By carrying out the integration over $t$, we immediately get
\begin{equation}
\mathcal{B}_{M}T=-i(-1)^{n}\frac{\pi ^{2}}{\Gamma (n)}\left( \frac{M^{2}}{4}%
\right) ^{n-3}.
\end{equation}%
Afterwards we apply the continuum subtraction procedure using the
replacement
\begin{equation}
\left( M^{2}\right) ^{N}\rightarrow \frac{1}{\Gamma (N)}%
\int_{0}^{s_{0}}dse^{-s/M^{2}}s^{N-1},\ N>0,
\end{equation}%
where $s_{0}$ is the continuum threshold parameter. Then for the Borel
transformed and subtracted integral $T$, we get
\begin{equation}
\mathcal{B}_{M}T=-i(-1)^{n}\frac{\pi ^{2}4^{3-n}}{\Gamma (n)\Gamma (n-3)}%
\int_{0}^{s_{0}}dse^{-s/M^{2}}s^{N-1},
\end{equation}%
Calculations of the other terms $T[l,m]$ in QCD side of the sum rule can be
performed in a similar manner. In the general case of $T[l,m]$, for
continuum subtraction one should use more complicated formulas, full list of
which can be found in Ref.\ \cite{Agaev:2016srl}

\begin{widetext}

As a result, for $\Pi ^{\mathrm{OPE}}(M^{2}, s_{0})$ we get
\begin{equation}
\Pi ^{\mathrm{OPE}}(M^{2},s_{0})=\frac{1}{147\cdot 4^{12}5^{2}\pi ^{10}}%
\int_{0}^{s_{0}}dss^{7}e^{-s/M^{2}}+\sum_{\mathrm{i}=3}^{30}\Pi ^{\mathrm{(i)%
}}(M^{2},\ s_{0}),
\end{equation}%
where the first term is the perturbative contribution.

For the nonperturbative $\mathrm{(3)-(10)}$ dimensional terms we get:
\begin{eqnarray}
&&\Pi ^{\mathrm{(3)}}=\frac{7m_{s}\langle \overline{s}s\rangle }{%
3^{3}4^{9}5^{2}\pi ^{8}}\int_{0}^{s_{0}}dss^{5}e^{-s/M^{2}},\notag \\
&&\ \Pi ^{\mathrm{%
(4)}}=\frac{\langle g_{s}^{2}G^{2}\rangle }{2\cdot 4^{11}5^{2}\pi ^{10}}%
\int_{0}^{s_{0}}dss^{5}e^{-s/M^{2}},\ \ \notag \\
&&\Pi ^{\mathrm{(5)}}=\frac{m_{s}}{%
2\cdot 3^{3}4^{11}5^{2}\pi ^{8}}\left[ 2\langle \overline{q}g_{s}\sigma
Gq\rangle \right.   \notag \\
&&\left. \times \int_{0}^{s_{0}}dss^{4}\left[ 12783-2340\ln \left( s/\Lambda
^{2}\right) \right] e^{-s/M^{2}}+\langle \overline{s}g_{s}\sigma Gs\rangle
\int_{0}^{s_{0}}dss^{4}\left[ 346+120\ln \left( s/\Lambda ^{2}\right) \right]
e^{-s/M^{2}}\right] ,  \notag \\
&&\Pi ^{\mathrm{(6)}}=\frac{1}{2\cdot 3^{3}4^{7}5\pi ^{6}}\left[ 2\langle
\overline{q}q\rangle ^{2}+\langle \overline{s}s\rangle ^{2}+12\left(
2\langle \overline{s}s\rangle \langle \overline{q}q\rangle +\langle
\overline{q}q\rangle ^{2}\right) \right] \int_{0}^{s_{0}}dss^{4}e^{-s/M^{2}},
\notag \\
&&\Pi ^{\mathrm{(7)}}=\frac{m_{s}\langle g_{s}^{2}G^{2}\rangle }{%
3^{3}4^{11}\pi ^{8}}\left[ 5\langle \overline{q}q\rangle
\int_{0}^{s_{0}}dss^{3}\left[ -311+60\ln \left( s/\Lambda ^{2}\right) \right]
e^{-s/M^{2}}+\langle \overline{s}s\rangle \int_{0}^{s_{0}}dss^{3}\left[
865-36\ln \left( s/\Lambda ^{2}\right) \right] e^{-s/M^{2}}\right] ,  \notag
\\
&&\Pi ^{\mathrm{(8)}}=-\frac{1}{2\cdot 3^{4}4^{15}\pi ^{10}}\left[
1429\langle g_{s}^{2}G^{2}\rangle^2 +21576\pi ^{4}m_{0}^{2}\left( 181\langle
\overline{q}q\rangle ^{2}+5\langle \overline{s}s\rangle ^{2}+342\langle
\overline{q}q\rangle \langle \overline{s}s\rangle \right) \right]
\int_{0}^{s_{0}}dss^{3}e^{-s/M^{2}},  \notag \\
&&\Pi ^{\mathrm{(9)}}=\frac{m_{s}}{2\cdot 3^{4}4^{12}\pi ^{8}}\langle
g_{s}^{2}G^{2}\rangle m_{0}^{2}\int_{0}^{s_{0}}dss^{2}\left[ -77464\langle
\overline{q}q\rangle +13922\langle \overline{s}s\rangle -\left( 27320\langle
\overline{q}q\rangle -748\langle \overline{s}s\rangle \right) \ln \left(
s/\Lambda ^{2}\right) \right] e^{-s/M^{2}}  \notag \\
&&-\frac{m_{s}}{2\cdot 3^{3}4^{4}\pi ^{4}}\left[ -12\langle \overline{s}%
s\rangle ^{2}\langle \overline{q}q\rangle +14\langle \overline{s}s\rangle
\langle \overline{q}q\rangle ^{2}+24\langle \overline{q}q\rangle ^{3}\right]
\int_{0}^{s_{0}}dss^{2}e^{-s/M^{2}},  \notag \\
&&\Pi ^{\mathrm{(10)}}=\frac{\langle g_{s}^{2}G^{2}\rangle }{3^{3}4^{9}\pi
^{6}}\left[ 386\langle \overline{q}q\rangle ^{2}+34\langle \overline{s}%
s\rangle ^{2}+636\langle \overline{s}s\rangle \langle \overline{q}q\rangle %
\right] \int_{0}^{s_{0}}dss^{2}e^{-s/M^{2}}-\frac{m_{0}^{4}}{3\cdot 4^{9}\pi
^{6}}\left[ 27\langle \overline{q}q\rangle ^{2}+\langle \overline{s}s\rangle
^{2}\right] \int_{0}^{s_{0}}dss^{2}e^{-s/M^{2}}.  \notag \\
&&
\end{eqnarray}%
In $\Pi ^{\mathrm{(i)}}(M^{2},\ s_{0})$ we have assumed $\langle \overline{u}%
u\rangle =\langle \overline{d}d\rangle $ and denoted both of them as $%
\langle \overline{q}q\rangle $. We do not provide explicit expressions of
the terms $\mathrm{(i)>(10)}$.

\end{widetext}

\end{document}